\documentclass[11pt]{article}
\usepackage[left=2.5cm,right=2.5cm,top=2.5cm,bottom=2.8cm,a4paper]{geometry}
\usepackage[T1]{fontenc}

\usepackage[utf8]{inputenc}
\usepackage{authblk}
\usepackage[colorlinks,allcolors=blue,bookmarks=false,hypertexnames=true,breaklinks=true]{hyperref} 
\usepackage{filecontents}
\usepackage{notoccite}
\usepackage{breakcites}
\usepackage{amsmath}
\usepackage[strings]{underscore}

\usepackage[noibid,backend=biber,authordate-trad]{biblatex-chicago}
\bibliography{references}

\title{Making Sense of Consciousness as Integrated Information: \\ Evolution and Issues of IIT}

\author[1]{Kyumin Moon}
\author[2]{Hongju Pae}
\date{}
\affil[1]{Dept. of Philosophy, Seoul National University\protect\\
\textit{dkxnaks@snu.ac.kr}}
\affil[2]{Interdisciplinary Program in Cognitive Science, Seoul National University\protect\\ \textit{hjpae@snu.ac.kr}}

\begin{document}
\maketitle
\begin{abstract}
The purpose of this article is to provide an overall critical appraisal of Integrated Information Theory(IIT) of consciousness. We explore how it has evolved and what problems are involved in the theory. IIT is a hypothesis that consciousness can be explained in terms of integrated information. It argues that a number of fundamental properties of experience can be properly analyzed and explained by physical systems’ informational properties. Throughout the last decade, there have been many advances in IIT’s theoretical structure and mathematical model. In addition, like all hypotheses in the field of science of consciousness, IIT has given rise to several controversies and issues. In this context, a critical survey for IIT is urgently needed. To this end, we first introduce fundamental concepts of IIT and related issues. Thereafter, we discuss major transitions IIT has been through and point out related intra-model issues. Finally, in the last section, some theoretical, extra-model issues involved in IIT’s principles are presented. The article concludes by suggesting that, for the sake of future development, IIT should more seriously take metacognitive accessibility to experience.
\mbox{} \\ \\
\textbf{keywords:} \it{Integrated Information Theory, the science of consciousness, consciousness, experience, qualia, panpsychism, metacognition}
\end{abstract}
\mbox{} 

\section{Introduction}
\label{1}
Integrated Information Theory of consciousness (IIT) is a hypothesis that consciousness can be explained in terms of integrated information. Among other theories, IIT might be one of the most interesting—but also controversial—hypotheses in the field of science of consciousness. IIT is suggested as a principled theoretical framework with the explanatory and predictive power. It argues that a number of fundamental properties of experience can be properly analyzed and explained by physical systems’ informational properties. Further, IIT claims that this information-centered, mathematical framework would shed some light on many clinically difficult and ambiguous cases. For this unique approach, IIT has consistently attracted considerable scholarly attention from neuroscientists, information theorists, and even physicists for over a decade. Throughout this period, there have been many advances in IIT’s theoretical structure and mathematical model. Furthermore, like all hypotheses in the field of science of consciousness, IIT has also given rise to several controversies\parencites{Cerullo,Horgan2015}. In this context, there is an urgent need for a critical survey for IIT that would address the following questions: What are the essentials of IIT? What has been changed and what has remained? And what problems can emerge against it? In the present article, we attempt to provide an overall critical appraisal of IIT. 
\par For this critical review, we will introduce core concepts, major transitions of IIT and related \textit{intra-model} issues that are rooted in the mathematical formulation. Then, several theoretical \textit{extra-model} issues involved in IIT’s principles, which are not directly due to the mathematical model, are outlined.\footnote{One of the reviewers has expressed some worries about the general structure of the manuscript of this article. The reviewer has pointed out that considering the critical motivations of this article, the manuscript contains too many technical reviews, which is not necessarily needed to justify theoretical criticisms developed in the latter of this article. The reviewer further advised that to clarify the common thread of the article, it would be better to focus on critical points and reduce technical details that are not directly relevant to the purpose of the article. Although we tried to shorten and revise the technical review, we believe at least some of the presentations of technical details are required, since they are necessary to understand the very points we made. In order to trace the version updates of IIT and reveal some problematic consequences of those updates, reviews of technical details are somewhat inevitable. Moreover, this article is intended to be an \textit{overall} critical appraisal, covering not only \textit{mathematical models} but also \textit{theoretical backgrounds} of IIT. Therefore, despite of lengthiness, we choose to stick to the initial structure of the article.} This article concludes by suggesting that, for the sake of future development, IIT should more seriously take metacognitive accessibility to experience. 

\section{Core Concepts of IIT}
\label{2}
Since IIT attempts to explain how conscious experience arises from physical substrates, there are several explanatory concepts describing this bottom-up process. While IIT has kept updating its version from 1.0 to 3.0\parencites{ton2001,ton2004,ton2008,ton2012,BT2008,BT2009,IIT3.0}, those core concepts remain to be fundamentals of the theory throughout all versions. Yet, despite their significant roles in the framework of IIT, the core concepts have not been clearly cashed out. To amend the situation, in what follows, we explain why those concepts are important in IIT and point out some related issues. While our characterization strongly reflects the view of the current version of IIT, providing a summary of IIT 3.0 is not the main concern in this section. Rather, following descriptions concern several central notions that persist regardless of versions. 

\subsection{Mechanisms, states, connections, and repertoires}
\label{2.1}
The central focus of IIT is on the physical substrates of experience and their causal structures. IIT analyzes candidate physical substrates of experience in a bottom-up manner; physical \textit{elements}, which can causally interact with each other, are under consideration. Any set of elements can be considered as a \textit{mechanism}. Furthermore, any set of mechanisms can be thought of as a higher-order mechanism or a \textit{system of mechanisms} (in short, \textit{system}). The system is composed of elements so that the system itself also can be a mechanism or a set of elements. On the other hand, causal structures of physical substrates are analyzed by two central notions of IIT; mechanisms, or systems, can be in a \textit{state}, which corresponds to outputs of their elements. For instance, if three elements—$A$, $B$, and $C$—with the binary output 1 or 0 compose a mechanism, and these element’s outputs are respectively $1$, $0$, and $0$, the state of the mechanism $ABC$ is represented as $100$\parencite[see][Figure 1A]{IIT3.0}. Further, such mechanism in a state can have a \textit{connection}, which corresponds to a set of causal connections among elements of the mechanism\parencite{BT2009}.\footnote{In \parencite{BT2009}, the term ‘submechanism’ or ‘mechanism’ was originally used to refer to sets or subsets of causal connections among elements. However, this use of the term causes a serious confusion, as ‘mechanism’ is also used in IIT to refer to sets or subsets of elements that causally interact. In order to avoid possible confusions, in the present paper, we use the term ‘connection’ instead of ‘submechanism’ or ’mechanism’.}  For example, if causal connections $c^1$, $c^2$, $c^3$, and $c^4$ are given, there might be a set of connections, such as as $\{c^1, c^2\}$, $\{c^1, c^3\}$, $\{c^1, c^2, c^3\}$ or $\{c^1, c^2, c^3, c^4\}$, etc. Any causal relationships could be characterized as a connection, such as synapses between neurons, which could be ideally represented as logic gates with simple computational functions. 
\par From states and connections of the mechanism, one can have \textit{repertoires}. A repertoire is defined as a \textit{probability distribution} to possible states of the mechanism. In IIT, the causal structure of the mechanism must be known \textit{a priori}.\footnote{This \textit{a priori} known causal structure can be mathematically described by the backward and forward Transition Probability Matrix (TPM) of the set of elements under consideration. While this requirement of \textit{a priori} given causal structure is usually not explicitly presented in literature, it is important and intrinsic to IIT. Thanks for the reviewer who clarified this point.} When the state and connection of a mechanism are given at time $t$, one can infer which past or future states of which mechanism—including the mechanism \textit{itself}—could be causes or effects of the given state of the mechanism, and how much probabilities would be distributed to each possible cause or future effect states. Therefore, these probability distributions are probabilistic expressions of how the mechanism’s particular state could cause or be caused by a certain mechanism’s past or future states. In this sense, the mechanism in the state \textit{specifies} repertoires, or its possible causes and effects. 
\par The notions of mechanisms, states, connections, and repertoires are the very fundamentals in IIT. Without these concepts, calculating information from a mechanism’s causal structure is not possible. As explained above, repertoires are derived from states and connections of the mechanism. Furthermore, as we will see in Section \hyperref[2.2]{2.2}, the very concept of information is formally defined by repertoires and related notions. The concepts of mechanisms, states, connections, and repertoires tie causation and information together and enable us to calculate how much information is generated from the causal structure of the mechanism. In part, this is the reason why they survived several updates so far. 
\par These notions also provide IIT with a quite liberal view about possible physical substrates of consciousness. None of these notions tells about what kind of materials should be considered as a candidate for the physical base of experience. Therefore, when something has its state and connection and specifies repertories, it can be at least considered regarding if it produces experience. Given that mechanisms or systems in a state are not limited to biological substrates, chemical structures such as silicon chips can be legitimate candidates for the physical base of consciousness. Thus, under the framework of IIT, the question “Is this cellular phone conscious?” is not a category-mistaken question that should be \textit{a priori} rejected. As far as the cellular phone can be considered as a “system of mechanisms in a state”, we can at least consider the possibility of its consciousness. In principle, anything that has its states and connections can be a mechanism, and any mechanism can be a possible candidate for a conscious mechanism\parencite{tonkoch2015}. 
\par However, such liberalism comes at a price. While the notions of mechanism, states, connections, etc., do not limit the \textit{kinds} of physical substrates of experience, they do not limit the \textit{levels} of physical substrates either. Said differently, those basic concepts do not identify in which \textit{spatio-temporal grains} we should find physical substrates of consciousness. Technically, there are elements, mechanisms, systems, state, and connections at each level of the grain; basic particles in microphysical interaction compose quantum mechanisms in a quantum state. Molecules in chemical bonding constitute chemical mechanisms in a chemical state. Neurons connected with synapses make neuronal mechanisms in a neuronal state. Among these levels, which mechanism should be taken as the origin of consciousness? The same applies to macro-levels. For example, in IIT, there appears to be no principled reason not to take China as a single mechanism in a state, composed of causally interacting Chinese people\parencite{sch2012}. Indeed, the problem of finding a proper spatio-temporal grain of consciousness has been admitted by IIT theorists themselves\parencites{ton2008,ton2012,IIT3.0}. We think that the problem already lies in the center of the basic notions of IIT, leaving theoretical loose ends.
\par It can be argued that the problem of spatio-temporal grains has already addressed in the current version of IIT.\footnote{One of the reviewers reminded us that IIT theorists already dealt with this problem of proper spatio-temporal granularity in \parencite{tonkoch2016}.} Applying the exclusion postulate introduced in IIT 3.0, proponents of the theory may argue that the appropriate spatio-temporal grains are ones that have \textit{maximum intrinsic cause-effect power}, which is quantified by the highest value of \textit{integrated conceptual information}\parencites{ton2012,tonkoch2016}. It is nonetheless possible that there are multiple highest values of integrated conceptual information across different spatio-temporal levels. For instance, if a certain part of the cortico-thalamic system, which is at the macro-spatial level, and a few numbers of neurons in V1, which is at the micro-spatial level, produce the same highest values of integrated information at the same time, then which level should be chosen as the level where experience arises? Unless principled solution being suggested, the problem of proper spatio-temporal grains would remain. 

\subsection{Intrinsic and causal information}
\label{2.2}
According to IIT, an amount of information generated by a mechanism is calculated from repertoires. This calculation is performed by measuring the distance between the unconstrained and constrained repertories. For the past or future state, IIT supposes the unconstrained repertoire as a probabilistic base. Given the system's causal structure, the repertoire is unconstrained in that such uncertainty is not constrained yet by the given state of the mechanism. Using Bayes’ theorem, one can infer the constrained repertoire from the given state of the mechanism. It is this distance between unconstrained and constrained repertoires that is defined as information throughout all versions of IIT.\footnote{For further detail on this calculation, see Section \hyperref[3.2]{3.2} and \parencite{IIT3.0}.}
\par The crucial point here is that those repertoires involved in information should be inferred from mechanisms \textit{within} a considered system.\footnote{As one of the reviewers clearly pointed out our inconsistent terminology in the original manuscript, we could correct this paragraph. The notion of “considered system” or “system under consideration” is explicitly introduced as \textit{candidate set} in IIT 3.0. A candidate set is \textit{a set of elements under consideration}; if certain elements are not included in the candidate set, those elements are considered as external noise, even if they still are part of the whole system. For further details on candidate set, see \parencite{IIT3.0}, Figure 1A.} To calculate repertoires specified by the mechanism in the state, one must consider past or future states of mechanisms within the system under consideration. No mechanism \textit{outside} of the considered system should be taken into account. For example, to calculate the amount of information generated by the mechanism mentioned in Section \hyperref[2.1]{2.1}, $ABC$ in $100$, one should consider mechanisms only within a considered system; suppose that with the mechanism of $ABC$, an element $D$ constitutes a certain system under consideration. Other elements, such as $E$ and $F$, are out of the considered system. Then, according to IIT, $ABC$ in $100$ cannot specify repertoires of mechanisms such as $E$, $EF$, or even $AE$, $AF$, $ABE$, $ABF$, $ABCE$, $ABCF$. It only specifies repertories of mechanisms $A$, $B$, $C$, $AB$, $AC$, $BC$, $ABC$, or $AD$, $ABD$, $ABCD$, $BD$, $ACD$, $CD$ and so on. Those repertoires would represent possible causes or effects of $ABC$’s being in $100$ that are in the considered system with their probabilities. In short, mechanisms in a certain system under consideration only can specify repertoires of mechanisms within that system. In this specific sense, in IIT, repertoires specified by the mechanism in a state express \textit{intrinsic causal power} of the mechanism. As repertoires represent the intrinsic causal power of the mechanism, information in IIT is essentially \textit{intrinsic} and \textit{causal}. Information generated by the mechanism is measured as the distance between repertoires. Of note, these repertoires involve nothing external to the system. They solely depend on possible causes or effects within the system. Therefore, information is intrinsic to the system in that it does not require anything external to the system. In addition, information has nothing to do with input/output signals that can be detected only by the external observer. Rather, it is about causes and effects that can be detected only from \textit{the system’s own intrinsic perspective}\parencites{ton2008,ton2012,IIT3.0}. Moreover, given that repertoires specify possible causes or effects and their probabilities, information produced by the mechanism is causal. This is why IIT repeatedly emphasizes the notion of information as \textit{“differences that make a difference”}\parencite{Bateson}. In IIT, for instance, the mechanism in a state specifies which past states of a certain mechanism (“differences”) would likely to cause the mechanism’s being in that state (“a difference”). This further implies that only something that can be \textit{selectively} caused or cause can produce information. This intrinsic and causal notion of information is the hallmark of IIT, which distinguishes IIT from other information theories: anything informative has an intrinsic causal power, and anything intrinsically causal has information. This intrinsic and causal nature of information is directly inherited by the most central concept in IIT, integrated information. Although integrated information is defined in a sophisticated manner, in so far as it is information, it also should be intrinsic and causal. The intrinsic and causal information is fundamental to IIT in that it determines what kind of information the theory deals with.
\par Although intrinsic and causal information constitutes one of the unique aspects of the theory, it also brings some problems concerning the \textit{function of consciousness}. Simply put, intrinsic and causal information does not involve anything outside of the system. By definition, integrated information has nothing to do with \textit{causal inputs/outputs} of the system either. This intrinsicness renders integrated information irrelevant to the functions of the system. In fact, as we will see in Section \hyperref[4.2]{4.2}., IIT theoretically designs functional zombie systems, which share all the input-output relations with systems with highly integrated information. This suggests that integrated information is nearly irrelevant to functions of the system; integrating information has no necessary bearing on the system’s functioning\parencite{sch2014}. In the sections below, we will see that IIT identifies integrated information and consciousness. If so, integrated information’s functional irrelevancy would be directly transferred to consciousness. For instance, IIT implies that, at least in principle, there can be perfect functional equivalents of us that are unconscious. It is at least theoretically possible that we do whatever we are doing without consciousness. Such state makes the ‘use’ of consciousness as mysterious. Moreover, \textit{adaptive benefits} of having experience also become doubtable; whatever adaptive function experience provides, there is always a possible scenario that it might have been evolved without experience. This risk of functional irrelevancy of experience has been already rooted in the intrinsic nature of information in IIT.

\subsection{Integrated information and complex}
\label{2.3}
The notion of \textit{integration} first stems from the phenomenological aspects of experience: “[p]he\-no\-me\-no\-lo\-gi\-cally, every experience is an integrated whole, one that means what it means by virtue of being one, and which is experienced form a single point of view”\parencite[p.295]{ton2012}. To be a physical underpinning of such integrated, unified experience, what should a mechanism be like? Here, IIT suggests one of its thought experiments: let’s compare a highly informative, but unconscious mechanism and a conscious mechanism. For example, what is the difference between a conscious brain and an unconscious digital camera that consists of thousands of photodiodes? According to the IIT, the most significant difference is that while the former is causally integrated, the latter is not\parencite{ton2012}. Causal interactions within the brain are so highly integrated with each other that, once they are fragmented, the whole brain’s performance might break down. This thought experiment on the camera model suggests that producing information is not sufficient for a mechanism to generate consciousness. Even if the mechanism is equipped with complicated connections and distinguishes vast repertoires, if its elements are not integrated into a single mechanism, the mechanism cannot give rise to experience. 
\par As a mechanism with a causal structure produces intrinsic information, one with integrated causal structure generates integrated information. The integrated information is integrated in the sense that, as a whole, the mechanism generates more information than the sum of its parts. Said differently, it is information produced only from the mechanism as a whole. By definition, the integrated information of the system is irreducible to its parts. Therefore, according to IIT, the amount of integrated information generated by the mechanism is calculated by partitioning the system by disconnecting the connections between the mechanisms. That is, if the information disappears by partitioning, it would be the information generated by the mechanism as a whole, not by individual parts. The informational difference between the mechanism as a whole and the system’s partitions’ mechanism is defined as integrated information.\footnote{This idea of integration might be closely related to the notion of synergy information proposed by Virgil and Koch\parencite{virgil}.} Nonetheless, considering that there are many possible ways of how the mechanism is partitioned, it becomes crucial to decide which partition should be used in calculating integrated information. IIT chooses the partition which causes the least loss of information, which is called \textit{minimum information partition}(MIP). Finally, depending on the level of calculation, the calculated values of integrated information are represented as $\Phi$ or $\phi$.
\par Based on the integrated information, a \textit{complex} is defined: roughly put, parts of the system producing integrated information can be considered as complexes.\footnote{Technically, how complexes are defined depends on which version of IIT is taken. As one reviewer noted, a complex is defined as a set of elements which produces a local maximum of integrated conceptual information on a system level, which quantified by $\Phi\textsuperscript{max}$ in IIT 3.0. While this is true, strictly speaking, notions such as integrated conceptual information, local maxima, and mechanism-system distinction were explicitly introduced since IIT 3.0. One cannot find any of these before IIT 3.0. In the IIT 2.0, complexes are defined differently, as sets of elements that produce integrated information. As we have noted at the beginning of Section \hyperref[2]{2}, our purpose is not briefly presenting the current version of IIT. (If it was, we would not cite any of the literature based on IIT 2.0 framework, including \parencite{ton2008} or \parencite{BT2008}, \parencite{BT2009}. The focus is on the core concepts that play essential roles \textit{throughout} all versions of IIT. Thus, until the end of Section 2, we temporarily choose to ignore conceptual differences among various versions and use some terms very loosely. In Section 2, for instance, “integrated information” covers integrated information in IIT 2.0 as well as integrated conceptual information and maximally integrated conceptual information in IIT 3.0. Accordingly, $\Phi$ can refer both $\Phi$ and $\Phi\textsuperscript{max}$.\label{fn8}} As we will see in Section \hyperref[2.4]{2.4}, IIT posits the identity between consciousness and integrated information; complexes in the system directly contribute to consciousness by integrating information. Technically, only complexes should be regarded as physical substrates of conscious experience, and they deserve to be called a ‘locus’ of consciousness. Despite a significant change concerning whether the overlapping or inclusion among complexes is possible, IIT maintains that a system can be condensed into multiple complexes. Finding such complexes in the system is the main focus of IIT on defining the local and temporal origin of consciousness. 
\par It should be emphasized that the notions of integrated information and complex provide possible explanations for some fundamental properties of experience. In calculating integrated information, nothing outside of the complex matters. How much information is generated by the complex, or how much information is lost by MIP, is purely intrinsic to the complex. This intrinsicness of integrated information accounts for why experience is essentially intrinsic. In other words, experience is integrated information, and integrated information is intrinsic. Therefore, experience is intrinsic. This characteristic of experience can be also noted as \textit{privacy}: “Since integrated information is generated within a complex and not outside its boundaries, the experience is necessarily private and related to a single point of view or perspective”\parencite[p.295]{ton2008}. Appealing to the central concepts such as integrated information and complex, IIT appears to open up the prospects of making sense of essential features of experience.
\par While integrated information is undeniably the key concept in the theory, it also creates a problem which renders the application of IIT to real systems practically intractable. As specified above, calculating integrated information involves finding MIP, which requires creating all possible partitions and measuring all the informational difference between the non-partitioned and the partitioned. With the growth of the number of the elements organizing the system, it becomes obvious that the amount of computation will dramatically increase. Consequently, one faces a serious \textit{combinatorial explosion} in finding MIP.\footnote{One reviewer mentioned that the levels of MIP need to be distinguished: the MIP on the level of small phi and the MIP on the level of conceptual information big Phi. It is of course true that these are two different forms of partitions that are respectively applied to the levels of mechanism and system. However, as clarified in footnote 8, such notions are restricted to the current version. They cannot be applied regardless of versions.} Due to this computational burden, applying IIT to neural substrates or artificial robots is currently infeasible. At the current stage of the theory, since direct empirical data supporting IIT are unavailable, researchers have tried to find efficient algorithms for finding MIP\parencites{OizKitzKanai2018,OizHid2018}, or to develop approximations or proxy measures of $\Phi$.\footnote{A large variety of modified $\Phi$ has been proposed as an estimation for $\Phi$. $\tilde{\Phi}_E$ is modulated by the Markovian discrete system and can be applied to continuous time series data\parencite{BS2011}, and $\Phi^{\ast}$ is modulated by substituting the notion of decoding perspective of information that facilitates the overall computation procedure\parencite{Oiz.decoding2016}. However, they do not contain the main theoretical updates of the IIT, such as cause-effect information and the distinction between $\phi$ and $\Phi$. For IIT 3.0, Marshall, Gomez-Raminez, and Tononi have proposed State Differentiation (SD) as a proxy measure of $\Phi$, which is much easier to draw out from experimental data than the original $\Phi$\parencite{TononiMar2016}. Still, it leaves the degree of integration being not properly measured. It is also insufficient to assume that SD functions are a complete form of measure in that it is applied to cellular animats; it is still analyzing the toy problem of the causal system. Recently, Tegmark has proposed several kinds of modified $\Phi$ through various definitions of informational distance and by normalizing integrated information using diverse techniques\parencite{teg2016}.} The absence of experimental validity is a decisive disadvantage for IIT to become a solid theory claim to the science of consciousness.  

\subsection{Identity between consciousness and integrated information}
\label{2.4}
As mentioned in Section \hyperref[2.3]{2.3}, IIT identifies consciousness with integrated information in the first place.  Particularly in IIT, levels of consciousness are identified with \textit{quantities} of integrated information, while \textit{qualities} of consciousness are identified with \textit{informational structures} derived from integrated information. From these identifications, IIT attempts to account for both how conscious a system is and how it feels. 
\par According to IIT, a level of consciousness is nothing but an amount of integrated information. Therefore, one can know how conscious the system is by calculating the amount of integrated information produced by that system.\footnote{Rich integration of neuronal connection is widely known as a major feature of the cerebral cortex, based on the anatomical structure of the brain. The fact that the brain cortex is constructed as a complicated neuronal network can explain how consciousness arises from such anatomical structure and why there exists such direct correlation between $\Phi$ and consciousness.} Consciousness is not all-or-nothing. Rather, as shown by the experience of falling asleep or that of anesthesia, consciousness is a matter of graded levels. IIT claims that a level of consciousness can be quantitatively measured by a value of integrated information, which is referred to $\Phi$. Since this ‘quantifying consciousness’ has drawn considerable scholarly attention, many IIT studies thus far have been dedicated to finding correlations between levels of consciousness and corresponding $\Phi$ values. Some of the results from analyzing EEG data and computer simulations suggest that $\Phi$ can be a reliable measure of consciousness\parencite{ton2008}. Indeed, the idea of the possibility to measure consciousness in a quantitative manner alludes to the science of conscious experience. It is the identification of levels of consciousness and $\Phi$ values that make this idea possible. 
\par On the other hand, IIT claims that a quality of experience is just an informational structure assessed from integrated information. This informational structure can be represented as a \textit{geometrical shape} in the multidimensional space that “completely and univocally specifies the quality of experience”\parencite[p.224]{ton2008}. If so, how the system feels can be known by deriving what shape is represented by the integrated information it generates. Each version of IIT provides sophisticated procedures for illustrating shapes like polytopes on multidimensional space from given integrated information. These shapes specify informational relationships generated by complexes. It appears to be obvious that, if it is successful, this geometrical representation provides useful tools for analyzing qualities of experience. Qualities of experience have several fundamental aspects to be explained, such as similarities and differences, richness, heterogeneities, as well as compositional structures. Once qualities of experience are identified with shapes assessed from integrated information, those fundamental aspects can be explained by analyzing geometrical characteristics of shapes in the multidimensional space. This explanatory potential of geometrical approach might be the most distinctive part of IIT; “what it is like to experience” could be explained by the “geometry of integrated information”\parencite{BT2009}. 
\par The identity between consciousness and integrated information has further implications. First, in clinical contexts, quantifying consciousness by $\Phi$ value might play a significant role in treating pathological cases of patients in coma or vegetative state. As the locked-in syndrome case suggests, how to judge whether or not one is conscious has been an extremely controversial issue. However, if $\Phi$ is indeed a level of consciousness, we have a simple answer: a patient is conscious only when his or her brain generates non-zero $\Phi$. This answer immediately leads to a liberal approach to the consciousness of non-humans. For animal consciousness, animals can be conscious if they generate integrated information at all. The same applies to artificial consciousness, as there is no reason to \textit{a priori} exclude the possibility that artificial intelligence can be conscious. The only thing that matters upon consciousness is whether or not the candidate system produces non-zero $\Phi$\parencite{tonkoch2015}. Second, the idea of qualities of experience as geometrical shapes appears to entail that experience is \textit{substrate-independent}. According to IIT’s geometrical approach, systems with different states and connections can produce the same informational structures\parencites{BT2008,BT2009}. From the assumption that qualities of experience are nothing but informational structure geometrically depicted from integrated information, it follows that qualities of experience and their physical substrates can come apart. This substrate-independence might account for, at least partially, why consciousness seems non-physical\parencite{teg2017}. In this respect, many theoretically and practically promising predictions and explanations come from the identification of consciousness and integrated information. 
\par However, the notion of consciousness as integrated information also raises some perplexing issues. Since the theory identifies $\Phi$ into a level of consciousness, IIT must ascribe experience to seemingly unconscious systems. Surprisingly, according to IIT, a photodiode with the binary states on/off is minimally conscious, because it produces non-zero $\Phi$.\footnote{Again, the reason why a photodiode should be treated as minimally conscious can be different according to versions of IIT. In IIT 2.0, the photodiode is minimally conscious, since it produces one bit of integrated information. However, in IIT 3.0, it is so because it generates non-zero integrated conceptual information. Although this distinction is important, for the purpose of Section \hyperref[2]{2}, we do not deal with differences among versions. See footnote \hyperref[fn8]{8}.} What is more, a lattice structure composed of a single kind of logic gates can be highly conscious, even more conscious than that of a human. In this way, IIT predicts that the functional zombie is empirically possible in principle\parencite{IIT3.0}; even if the two systems are functionally equivalent, there can be a situation when one generates $\Phi$, but the other does not\parencite[see][Figure 21]{IIT3.0}. The theoretical development of IIT is not without a sense of irony; core concepts promising ample explanatory and predictive potentials are also bringing counterintuitive and problematic defects to the theory. In the remainder of this paper, these intriguing issues will be analyzed in further detail. 

\section{Major Transitions in IIT}
\label{3}
While the core ideas of IIT are more or less preserved, it has undergone several significant revisions through the updates from its prototype to the very latest version. These revisions made the framework of IIT more theoretically and technically articulated. Not only its theoretical structure but also the details of its mathematical model have been changed. Therefore, understanding how IIT has acquired its current form requires a deeper analysis. Axioms, postulates, and the notion of purviews have been introduced, the concept of information has been revised, the distance matrix for repertoires has been changed, and levels of information integration have been divided. In what follows, we explain what these changes are about and how they affect IIT. 

\subsection{From thought experiments to systematic formulation: Phenomenological axioms and ontological postulates}
\label{3.1}
As occasionally mentioned in the literature, what consciousness is and what a physical system should be to generate consciousness have never been explicitly described until IIT has developed into its latest version. The fundamental properties of experience were taken for granted by appealing to phenomenology, and the required properties for physical systems to produce experience were merely motivated or suggested by a number of thought experiments. The photodiode and camera thought experiments were introduced from the early version of IIT\parencites{ton2004,ton2008}, and the Internet thought experiment was added during the updates to IIT 3.0\parencite{ton2012}. Based on the idea that conscious experience is specific in a particular way, the photodiode thought experiment motivates that physical systems must specify its possible causes or effects to generate consciousness. The digital camera thought experiment suggests that physical systems must be causally integrated since it appears that experience is unified and integrated. By contrasting information of the Internet and that of experience, the Internet thought experiment speculates that information produced by physical systems must be maximally integrated. While these thought experiments are interesting in themselves and might be helpful to understand the motivations behind the theory, they never clearly argued for or even specified the fundamental features of the essential properties of experience. 
\par In IIT 3.0, the situation has changed. Now, the fundamental properties of consciousness and requirement for physical systems are explicated and posited in the very beginning of the theoretical formulation. First and foremost, \textit{phenomenological axioms} are introduced; these axioms are phenomenological in the sense that they are all concerned with the fundamental properties of experience. Each of five axioms corresponds to each of the essential properties of experience. The \textit{existence} axiom states that consciousness exists. The \textit{composition} axiom says that it is compositional. The \textit{information} axiom states that it is informative. The \textit{integration} axiom claims that it is integrated. Finally, the \textit{exclusion} axiom says that one consciousness excludes another consciousness\parencites{ton2012,IIT3.0}. These properties mentioned in the axioms are supposed to be fundamental, as \textit{any} experience must have them. 
\par Next, corresponding to the phenomenological axioms, \textit{ontological postulates} are posited: these postulates are ontological in that they prescribe what mechanisms should generate consciousness. There are five postulates which lay parallel to each axiom. The \textit{existence} postulate says mechanisms in a state must exist. The \textit{composition} postulate says that mechanisms must be structured. The \textit{information} postulate claims that mechanisms must produce information by specifying selective possible causes and effects within the system. The \textit{integration} postulate states that mechanisms must integrate information. Finally, the \textit{exclusion} postulate says that mechanisms must generate only the maximally integrated information\parencites{ton2012,IIT3.0,tonkoch2015}. As in phenomenological axioms, properties mentioned in ontological postulates are essential and necessary for \textit{every} physical mechanism to generate consciousness. Moreover, the latter three postulates—information, integration, and exclusion—are applied to the two different levels of calculation; \textit{mechanisms} and \textit{systems of mechanisms}. Contents of postulates vary through the system depending on which level they are applied to.\footnote{For further details on ontological postulates, see \parencite{IIT3.0}.}
\par In virtue of these axioms and postulates, IIT becomes a top-down and theory-driven approach, rather than as a bottom-up and experiment-driven approach to consciousness; the set of axioms and postulates comes first and later comes the mathematical model. Empirical experiments can be designed and conducted only under the models and theories. Consequently, by declaring its axioms and postulates, IIT can clarify both its theoretical framework on modeling the consciousness and the experiments.
\par However, while clarification is one thing, justification is another thing. The introduction of the axioms and postulates raises a number of questions. First, on what ground must phenomenological axioms be accepted? That is, why should those axioms be considered axiomatic? The axioms themselves appear to be based on \textit{phenomenological intuitions} or \textit{introspection}. The axioms are “assumed to be self-evident from the intrinsic perspective of a conscious entity”\parencite[Supplementary 1, p.1]{IIT3.0}. However, what if such intuitions or introspections are wrong? Moreover, how can each ontological postulate follow from each phenomenological axiom? Though the postulates are strictly parallel to axioms, there seems to be an unbridged gap between them. For instance, it is not clear how we can draw the information postulate; it is not clear if the mechanisms should specify selective causes and effects within the system from the information axiom, which states that consciousness is informative. Therefore, the rationales for positing phenomenological axioms and ontological postulates remain controversial. Recently, Bayne addresses precisely the axiomatic foundations of IIT\parencite{Bayne}. Bayne argues that some of the phenomenological axioms are not self-evident, and others seem to be self-evident but fail to practically or theoretically constrain the theory of consciousness at all. He suggests that IIT would be on firmer ground if it adopts what he calls ‘natural kind approach’\parencite{Bayne}. While the verdict may still be out, it appears that the axiomatic approach and seemingly following postulates are not secured as they seem.\footnote{We the authors gratefully thank the reviewer who recommended the recent literature. It was truly helpful to strengthen our argument.}

\subsection{From effective information to cause-effect information: New information and its metric}
\label{3.2}
The revision of the notion of information might be one of the most significant developments during the updates of IIT. In the early versions of the theory, information generated by a system was defined as \textit{effective information}($ei$)\parencite{ton2008}. There were two kinds of repertoires: a potential repertoire, which is a probability distribution of the past states when no current state of the system is known, and an actual repertoire, which is a probability distribution of the past states when a particular current state of the system is known. One can measure the distance between the potential and the actual repertoires by applying \textit{Kullback-Leibler Divergence}(KLD), and such distance could be thought of a sort of \textit{relative entropy}. This distance or relative entropy directly equals to the effective information. In IIT 3.0, however, a new form of information is introduced: \textit{cause-effect information}($cei$)\parencites{ton2012,IIT3.0}. The cause-effect information differs from effective information in many important aspects. 
\par First, unlike $ei$, $cei$ involves the system’s past and future\parencite{ton2012,IIT3.0}. In calculating $ei$, potential and actual repertoires are only of the past states of the system. In calculating $cei$, however, repertoires concern \textit{both} the past \textit{and} future states of the system. These repertoires can be thought of as probabilistic expressions of how the current state of the mechanism would be caused by the past states of the system \textit{and} how it would cause the future states of the system. On one hand, there are \textit{unconstrained past repertoire} and \textit{cause repertoire}. The former is a probability distribution of the past states of a certain mechanism of the system when the current state of the given mechanism is not known. It always produces maximum entropy of past states. The latter is defined as a probability distribution of the past states of a certain mechanism of the system when the current state of the given mechanism is known. On the other hand, there are \textit{unconstrained future repertoire} and \textit{effect repertoire}. The former is a probability distribution of the future states of a certain mechanism when the current state of the given mechanism is perturbed in every possible way. The latter repertoire is a probability distribution of the future states of a certain mechanism of the system when the current state of the given mechanism is known. The current state of the given mechanism specifies cause and effect repertoires of a certain mechanism of the system. It should be noted that unconstrained past and future repertoires and cause and effect repertoires are calculated \textit{independently} of each other. Therefore, repertoires should be calculated \textit{twice} in calculating $cei$; therefore, while calculating ei requires only two repertoires, four repertoires are required to calculate $cei$. 
\par Second, while $ei$ concerns only the states of the given system itself, $cei$ can involve the past/future states of mechanisms \textit{other than} the given mechanism\parencites{ton2012,IIT3.0}. For $ei$, only the past states of the same system should be taken into account, as potential and cause repertoires are defined as probability distributions of the past states of the system itself and nothing else. However, in calculating $cei$ produced by the mechanism, not only the past/future states of the mechanism itself but also those of other mechanisms of the system can be considered. Said differently, the repertoires required to calculate $cei$ of the given mechanism are not restricted to the same mechanism. For example, if the whole system is composed of elements $A$, $B$, and $C$ with binary outputs 1 and 0, and the selected mechanism is $A$, $A$’s current state 1 can specify the cause repertoire of the past states of any mechanism, including $B$, $BC$, $AC$, $ABC$, or even $A$ itself. Similarly, it can also specify the past repertoire of the future state of any mechanism\parencite[see][Figure 4]{IIT3.0}. Thus, to calculate $cei$ generated by $A$ in 1, one must first decide which mechanism to be paired with $A$. In principle, any mechanism of the system, which could be represented as the power set of the total elements of the system, can be paired with $AB$. In IIT 3.0, this idea of pairing is introduced as purview. When $A$ in 1 is paired with $ABC$ and the cause repertoire is calculated, the purview of $A$ is represented as $A^c/ABC^p$. If it is paired with $ABC$ and the effect repertoire is calculated, the purview of $A$ is represented as $AB^c/ABC^f$\parencite[see also][Figure 4]{IIT3.0}. Once the purview is fixed, other elements outside the purview remain unconstrained and do not affect cause and effect repertoires. However, calculating unconstrained repertoire does not require a specific purview, because there is no difference on unconstrained repertoires among different current purviews. By discriminating the mechanism’s purview, the causal analysis could be extended to every mechanism of the system. 
\par Third, whereas $ei$ is measured by KLD, $cei$ is measured by a different metric. As explained above, to calculate $ei$, we must measure the distance between the potential and the actual repertoires. In the earlier versions, it was KLD which was used to measure the distance. KLD is the most intuitive index for measuring the reduction of entropy, which directly relates to the quantity of information generated in terms of relative entropy. Since entropy and information were regarded as symmetrical, KLD was chosen for the scale of distance during the early versions of IIT. However, technically, KLD should not be considered as a proper metric, since it is not symmetric, does not obey triangular inequality, and is unbounded. In addition, non-compensated KLD measures only the reduction of uncertainty and does not account for the difference between states, which appears to be crucial in calculating information. For these reasons, another measure should be introduced as a new scale\parencite[see][Supplementary 2]{IIT3.0}. 
\par Therefore, from IIT 3.0, \textit{Earth Mover’s Distance}(EMD) is used to measure the distance between repertoires. This is also known as Wasserstein distance, which is the distance function defined by the minimum cost of redistributing the “dirt piles” to the location elsewhere\parencite{IIT3.0}. Given that distributed probabilities can be thought of as “dirt piles”, one can think of a distance between two repertoires as the minimum cost of distributing “dirt piles”. In IIT, there are in fact two kinds of EMD. First, a general EMD is applied in calculating cause-effect and integrated information on the level of mechanisms. Second, an extended EMD is used to calculate that on the level of the systems of mechanisms. Nevertheless, in both cases, the point of using EMD remains the same. By using EMD, not only the reduction of entropy but also the difference between states is taken into account in calculating information. From IIT 3.0, the quantitative value of information is not represented by bit, since the unit of the distance measured by EMD is not a bit. In sum, EMD appears to be a more appropriate metric for IIT than KLD. Based on EMD, the distance between unconstrained past repertoire and cause repertoires is defined as the \textit{cause information}($ci$). The cause information illustrates possible causes of the mechanism’s current state when a purview of the mechanism is fixed. Similarly, the distance between unconstrained future repertoire and effect repertoire is defined as the \textit{effect information}($ei$). This implies that effect information signifies possible effects of the mechanism’s current state when a purview of the mechanism is fixed. In sum, effect information can be calculated from the distance between those repertoires and is quantified by EMD as same as cause repertoire.\footnote{After IIT 3.0, a large variety of distance functions, such as Hilbert-space distance and Shannon-Jensen distance, have been newly proposed as metrics of informational difference\parencite{teg2016}. It would be important to consider the characteristics of each measure in order to broaden the explanatory power of IIT.}
\par Finally, there is an informational principle that should be applied to cause-effect information. In the earlier versions of IIT, measuring the distance between repertoires was all that mattered. The measured distance was the amount of ei of the system. In IIT 3.0, however, there is more than just measuring the distance. As explained so far, there are two pairs of repertoires that lead to two kinds of information: cause and effect information. Then, which information should be accounted as the mechanism’s information? At this point, the \textit{Information Bottleneck Principle}(IBP) is introduced\parencite{IIT3.0}. IBP forces one to choose the \textit{minimum} of cause and effect information. The motivation behind IBP comes from intrinsic and causal notion of information: since information in IIT is supposed to be intrinsic to the system, the information that can be detected only by the external observer must be excluded. Suppose if the mechanism in a state only generates cause information, but no effect information. This implies that the mechanism being in such a state does not make any difference to the system. In that case, although the mechanism still belongs inside the system, it does not give any causal interaction among the system’s other mechanisms. Hence, such cause information produced purely by the mechanism cannot be detected from the intrinsic perspective of the system. The same holds when the mechanism in the state produces only effect information, but no cause information. This observation enforces IBP so that the smaller one between cause and effect information is taken as $cei$.
\par Since the concept of information is at the heart of the theory, the transition from ei to cei articulates the framework of IIT in a number of important aspects. By taking into account both the past and the future, the notion of information becomes more causal; it involves causes \textit{and} effects. The application of IBP makes it more intrinsic. Specifically, as we will see in Sections \hyperref[3.3]{3.3}-\hyperref[3.4]{3.4}, when it comes to integrated information on the level of mechanisms, considering all possible purviews of the given mechanism plays a crucial role in enforcing the exclusion postulate. This involves the central notions of IIT 3.0, including concepts, conceptual structure, and other related ideas. 
\par However, the technical complexity is the other side of the theoretical articulation. As mentioned above, the computational burden is doubled, since repertoires and information must be calculated twice. The multidimensional space for geometrical representation of concepts is also doubled\parencite[see][Figure 15]{IIT3.0}. Moreover, calculating integrated information of mechanisms becomes more computationally complicated, because it should concern possible purviews of the mechanism. To calculate integrated information of the mechanism, MIP should be found in each possible purview\parencite[see][Figure 8]{IIT3.0}. In this sense, it appears to be clear that the introduction of purview worsens the combinatorial explosion. In what follows, all these technical issues will be analyzed in further detail. 

\subsection{From one phi to two phis: the distinction between $\Phi$ and $\phi$}
\label{3.3}
Before IIT 3.0, there was only one kind of integrated information; all integrated information calculated from the system was noted as $\Phi$. Cause-effect repertoires were inferred from the mechanism, and it was all that mattered in calculating integrated information. However, from IIT 3.0, the distinction between the level of mechanism and that of systems of mechanisms has been introduced. According to this distinction of levels, a distinction between kinds of integrated information has been made\parencites{ton2012,IIT3.0}. On the one hand, there is \textit{integrated information} generated from mechanisms, which is indicated as $\phi$(small phi); on the other hand, there is \textit{integrated conceptual information} produced by systems of mechanisms, which is represented as $\Phi$(large phi). $\phi$ and $\Phi$ differ from each other both in their concepts and calculations. 
\par Integrated information $\phi$ succeeds the motivation \textit{“more than the sum of its parts”} from older versions of IIT and is still analyzed on mechanisms. According to IIT, if there is a difference between the sum of the cause-effect information created by the partition of mechanism and the cause-effect information generated by the unpartitioned mechanism, and this difference directly refers to the information which mechanism forms as a whole entity. Any possible subset of a mechanism which can make difference on repertoire could be a candidate for partition.\footnote{It is interesting that one of the variable subsets at a certain time might be empty as a result of a particular partition. Furthermore, partitioning can be thought of a method of making certain mechanisms causally inactive. This process is called ‘virtualizing the element’ or ‘injecting noise to the mechanism’. For more detailed analysis, see \parencite{KroOst2017}.} On the level of the mechanisms, $\phi$ can be measured by making a partition on a given purview; for example, the purview of $ABC$ in $100$ is defined over the past mechanisms in a state. As briefly explained in Section \hyperref[2.3]{2.3}, among all possible partitions, MIP is selected for the calculation of integrated cause information, $\phi\textsubscript{cause}$. The purview of $ABC$ also can be defined over the future mechanisms in a state. Applying the same procedure, integrated effect information, $\phi\textsubscript{effect}$, can be calculated. By IBP, one can have integrated information $\phi$. In this way, $\phi$ can be calculated respectively from every single purview available on a certain mechanism\parencite[see][Figure 8]{IIT3.0}.\footnote{For the details of these computational steps, see \parencite{IIT3.0}, Figure 6.}
\par However, since there can be many possible past/future purviews on a mechanism, one mechanism in a state can have a multiple possible $\phi$s. Here, one of the ontological postulates comes in: the \textit{exclusion postulate} states that, in order to contribute to experience, the mechanism must have only one set of possible causes and effects which is maximally irreducible, while all other sets should be excluded\parencites{ton2012,IIT3.0}. It means that only the cause-effect repertoire of the mechanism that provides the maximum value of $\phi$, $\phi\textsuperscript{max}$, should be taken. As $\phi$ is defined as the minimum of $\phi\textsubscript{cause}$ and $\phi\textsubscript{effect}$, in order to find $\phi\textsuperscript{max}$, one must find maximally irreducible cause repertoire that yields $\phi\textsubscript{cause}\textsuperscript{max}$ and maximally irreducible effect repertoire that provides $\phi\textsubscript{effect}\textsuperscript{max}$ first. Then, the minimum of $\phi\textsubscript{cause}\textsuperscript{max}$ and $\phi\textsubscript{effect}\textsuperscript{max}$ would be $\phi\textsuperscript{max}$. The maximally irreducible cause repertoire is called \textit{core cause}, the maximally irreducible effect repertoire \textit{core effect}. The pair of core cause and effect is noted as \textit{Maximally Irreducible Cause and Effect repertoire}(MICE). MICE or the mechanism which specifies MICE is called a \textit{core concept}, or just \textit{concept}.\footnote{IIT 3.0 show a serious inconsistency in using the term concept: on one hand, ‘concept’ seems to refer MICE, a maximally irreducible cause-effect repertoire. In \parencite{IIT3.0}, it is said: “the notion of a \textit{concept}: the maximally irreducible cause-effect repertoire of a mechanism”(p.3). On the other hand, it is also used to indicate mechanism which specifies the MICE: “If the MICE exists, the mechanism constitutes a \textit{concept}.”(p.3), “concept($\phi\textsuperscript{max}$): A mechanism that specifies a maximally irreducible cause-effect repertoire(MICE or quale “\textit{sensu stricto}”)”(p.5, Table 1), and “A mechanism that specifies a \textit{maximally irreducible cause and effect}(MICE) constitutes a \textit{concept}”(p.9). What is worse is that the term is described as denoting \textit{both}: “Concept: A set of elements within a system and the maximally irreducible cause-effect repertoire it specifies, with its associated value of integrated information($\phi\textsuperscript{max}$)”(p.5, Box 1). To avoid possible confusions, we choose the second use. In this article, the term concept will always refer to the mechanism specifying MICE.} In short, by the exclusion postulate, the highest value of $\phi$ should be chosen among all possible $\phi$s produced by the mechanism and is defined as $\phi\textsuperscript{max}$. Here, the mechanism that produces $\phi\textsuperscript{max}$ is should be regarded as the concept.
\par After finding concepts, one can calculate the amount of integrated conceptual information at the level of systems of mechanisms. Concepts can be illustrated as points in the multidimensional space called \textit{concept space}, and these points would make ‘constellations’ among the coordinate space. In IIT 3.0, the constellation of concepts is defined as \textit{conceptual structure}\parencite{IIT3.0}. As each mechanism specifies its own MICE, the system of mechanisms specifies its own conceptual structure in the concept space. From this conceptual structure, one can calculate \textit{Conceptual Information}(CI) generated by the system of mechanisms. As CI corresponds to the cause-effect information, it is quantified in a similar way; as there must be unconstrained past and future repertoires for calculating cause-effect information, there must be the \textit{“null” concepts} for calculating CI. “Null” concepts are the unconstrained past and future repertoires in which the state of the system of mechanism is undecided.\footnote{The “null” concepts are named so because they specify unconstrained past and future repertoires if considered as mechanisms; in other words, it is the concept that specifies nothing. Although it is perceived as only a superficial notion on designating unconstrained repertoire, however, it also can be illustrated in conceptual space along with other concepts\parencite[see][Figure 11]{IIT3.0}. By this way, “null” concept refers to the concept specifying its current purview of mechanism as an empty set.} By applying the extended version of EMD, it can be quantified how much CI is produced by the system of mechanisms.\footnote{Extended EMD differs from original EMD by its methods on calculation. The distance between cause-effect repertoire and unconstrained cause-effect repertoire is measured by EMD, and each EMD of cause and effect distributions are added up, then it is multiplied by $\phi\textsuperscript{max}$ of the concept, which functions as the weight of each concept. In short, extended EMD is used on the level of systems of mechanisms by multiplying $\phi\textsuperscript{max}$ as each distribution’s weight. Even if the details on the calculation vary, the fundamentals on calculating the distance, which is redistributing the probability distribution, do not change. For further detail on applying extended EMD, see \parencite{IIT3.0}, Supplementary 2.}
\par The calculation of integrated conceptual information $\Phi$ is also analogical to that of $\phi$. Once the purview of the system of mechanisms is given, MIP can be found by partitioning\footnote{The partitioning in the level of systems of mechanisms must be unidirectional\parencite{IIT3.0}. Unidirectional partitioning is done by virtualizing elements; when a mechanism is injected with noise, information disappears, as the mechanism gets considered as external noise and loses its intrinsic causal power. Unidirectional partitioning could be thought of as injecting noise between subsets for only to a certain direction of the connection. Thus, partitioning the direction of connection between subsets on the system level is analogous to virtualizing the elements on the mechanism level. For further detail on virtualizing the elements, see also \parencite{KroOst2017}.} the purview. By measuring the difference of CI between the unpartitioned and the partitioned, it can be calculated how much the integrated conceptual information is generated by the system of mechanisms. Again, as for $\phi$, there can be many possible $\Phi$s as all possible unidirectional partitions of the set of elements should be considered. Here, the exclusion postulate comes in again; it enforces only one complex among all other overlapping systems of mechanisms to contribute to consciousness. Thus, the one that generates the maximum of $\Phi$ should be chosen as \textit{$\Phi\textsubscript{max}$}. Finally, the conceptual structure that gives rise to $\Phi\textsubscript{max}$ is defined as \textit{Maximally Integrated Conceptual Structure}(MICS). The system of mechanisms that produces $\Phi\textsubscript{max}$ and so specifies MICS is defined as complex.\footnote{As a result, there comes an important change in defining the complex since IIT 3.0. In IIT 3.0, due to the exclusion postulate, complexes cannot be nested or overlap at all. In the earlier versions, however, since the exclusion axiom/postulate was not introduced yet, complexes could partially or wholly overlap. Meanwhile, there can be multiple complexes in one system. IIT predicts that one system can be condensed into \textit{several} complexes. The complex that has $\Phi\textsubscript{max}$ is called \textit{major complex}, and the complex which does not overlap, but has $\Phi$ smaller than $\Phi\textsubscript{max}$ is called \textit{minor complex}. Since they have their own $\Phi\textsubscript{max}$, they are considered as an individual complex. Minor complex can be thought of as a local maximum which implies ‘locally condensed minimal consciousness’\parencite[see][Figure 16]{IIT3.0}. There should be a single major complex in general situations, but there could be multiple major complexes according to circumstances. For example, split brain syndrome or dissociative disorders could be explained as clinical examples of the main complex being split into two or more. At the same time, minor complexes could be thought of as \textit{preconsciousness}; the constituent of consciousness which can contribute to the reaction of extrinsic inputs. Continuous flash suppression could also be explained through the function of the minor complex\parencite{IIT3.0}.} In IIT 3.0, such MICS generated by the complex is directly identified as the subjective experience. 
\par By introducing the distinction between $\phi$ and $\Phi$, now it is much logical to explain the generation of consciousness using integrated information in further detail. In IIT 3.0, MICE is called \textit{quale “sensu stricto”}, which means quale in the narrow sense. Since this sort of quale includes ‘redness of red’ or ‘painfulness of pain’, it can be considered to be quale in the philosophical debates. On the other hand, MICS is called \textit{quale “sensu lato”}, which means quale in a wide sense\parencite{IIT3.0}. As mentioned above, IIT equates experience with MICS. As mechanisms in the complex maximally integrate cause-effect information, concepts are generated, and we have qualia. As the complex maximally integrates CI, MICS is produced and we have an experience. Based on the distinction and the exclusion postulate, the notions of concepts and MICS can be defined. These central notions of IIT 3.0 enable one to explain how experience arises from its physical substrates in a bottom-up manner. All these articulated explanations essentially start from the distinction between $\phi$ and $\Phi$. 
\par However, this distinction between $\phi$ and $\Phi$ also raised several problems for IIT. First and foremost, computing $\Phi$ and the application to real systems became computationally intractable\parencite{IIT3.0}. Almost every aspect of calculation doubled: MIP had to be found twice; once at the level of mechanisms and twice at the level of systems of mechanisms. Owing to the distinction between $\phi$ and $\Phi$, it appears that the combinatorial explosion in IIT extremely deteriorated. In turn, such computational infeasibility rendered the empirical prospect of IIT more pessimistic.\footnote{One of the reviewers noted that the source of the combinatorial explosion is the combination principle. For example, as the calculation of $\phi$ needs be performed according to the combination principle, it must be carried out over all possible subsets of the candidate set and for each of those subsets over all possible purviews. Nevertheless, the main constraint is, that the number of unique bipartitions rises exponentially with the cardinality of the set\parencite[see][Appendix]{KroOst2017}.} At the cost of an articulated bottom-up explanation of experience, the theory had to face serious practical problems in retaining empirical validity. 
\par Another problem emerges from the exclusion postulate. As explained above, the exclusion postulate enforces that only the mechanisms which give rise to $\phi\textsubscript{max}$ or the systems of mechanisms which provide $\Phi\textsubscript{max}$ must be taken as the concept or complex. However, if there are several different MICEs that yield exactly the same $\Phi\textsubscript{max}$, then which repertoire or conceptual structure should be taken? Clearly, being as biggest does not involve being as unique. Nevertheless, the exclusion postulate says nothing about this problematic \textit{underdetermination of quale}\parencite{KroOst2017}. Moreover, as we have noted at the end of Section \hyperref[2.1]{2.1}, what if exactly the same $\Phi\textsubscript{max}$ are produced at the different spatio-temporal grains? When two equivalent $\Phi\textsubscript{max}$s are detected both from the level of neuronal units and from the level of cerebral lobes, the exclusion postulate cannot tell which level should be taken as the ‘locus’ of conscious experience. At least at the current stage of the theory, IIT does not have any theoretical resource to deal with such issues. 

\subsection{From vector geometry to point geometry: the geometry of integrated information}
\label{3.4}
IIT has always assessed integrated information in a geometrical manner. Nonetheless, several updates from IIT 1.0 to 3.0 brought a number of changes in the geometry of integrated information. Since IIT’s central notions, such as concepts and conceptual structure, are closely related to the geometry of integrated information, a deeper analysis of why and how the geometry has been revised would be needed.
\par In earlier versions of IIT, the space for representing informational structures was dubbed \textit{qualia space}, the multidimensional space which has its axes for each possible state of the system\parencites{ton2008,BT2008,BT2009}.\footnote{For example, when $n$ is the total number of the system’s elements and each element can have only two possible outputs, dimension of $2^n$ is required to constitute the qualia space which represents the system.} The geometrical shape expressing the informational structure was called \textit{quale}. The quale is constituted by q-arrows and points in qualia space: points represent actual repertoires specified by the system in a state when a certain connection—a set of causal connections—is added. Furthermore, q-arrows represent \textit{informational relationships} between each actual repertoire specified by the added connection. Thus, the point “at the bottom” of the quale is a potential repertoire specified by the system in a state, when no connection is added (the “null set”). On the other hand, the point “at the top” of the quale is the actual repertoire, when all connections are added (the “full set”). By adding each connection from the potential repertoire, it is possible to analyze how much $ei$ does the system gains by each connection. This can provide detail about which connection informationally contributes to the quale. All points connected by all q-arrows illustrate a geometrical figure like a “polytope”\parencite[see][Figure 3]{BT2009}. 
\par The major point of the geometry of the earlier versions of IIT is that q-arrows are represented as vectors. Interpreting q-arrows as vectors, one can find many properties of the informational relationships constituting the quale: the length of the q-arrow represents how much $ei$ is generated by adding a connection. For instance, the length of the q-arrow connecting “the bottom” and “the top” of the quale represents $ei$ of the system in a state. In addition, the direction of the q-arrow expresses the \textit{particular way} how adding a connection sharpens repertoires. One of the most interesting properties of q-arrows, however, would be \textit{entanglement}($\gamma$): when a q-arrow cannot be decomposed into an exact vector sum of its sub-q-arrow, then it is considered as tangled. When the q-arrow is tangled, it means that there is integrated information gained by adding up the corresponding connection. The way how the q-arrow is tangled can be measured by vector calculation. The difference between the length of the q-arrow and that of the vector sum of its sub-q-arrows is quantified by $\gamma$. In this sense, entanglement represents how much information a q-arrow generates above and beyond its components. In an earlier version of IIT, the q-arrow with $\gamma$>0 was defined as a \textit{concept}. Moreover, \textit{complexes} could be defined by comparing $\gamma$ of each concept; a concept with relatively high $\gamma$ was called as a mode. Before IIT 3.0, these articulated analyses were available from the vector analysis of q-arrows\parencites{ton2008,BT2008,BT2009}.
\par However, in IIT 3.0, such vector analysis is no longer available\parencites{ton2012,IIT3.0}. As explained in Section \hyperref[3.3]{3.3}, the concept cannot be defined as an entangled q-arrow. Rather, it is defined as MICE plotted as a point in the concept space. Instead of the null set, the current version posits the “null concept”, which is the unconstrained repertoire specified by the system of mechanisms when no mechanism of the system is given. As explained in Section \hyperref[3.3]{3.3}, one can calculate how much CI is generated by the mechanism by applying extended EMD. Thus, in concept space, the distance between two concepts does not capture how much $ei$ is generated by adding a connection to the system in a state. Rather, it captures how much CI is generated by adding a mechanism to the system of mechanisms. Due to these differences, adding connections, specifying informational relationships between repertoires, and analyzing q-arrows cannot be found in the current version of concept space. In sum, all analyses and notions grounded by vector calculus of q-arrows are not available in IIT 3.0. Prima facie, the geometry of integrated information appears to be simplified. 
\par The transition from effective information to cause-effect information also affects the geometry of IIT. The transition in the notion of information doubles the concepts and space. Since the theory was based on $ei$, there were only the past repertoires. Therefore, just one space was required to represent concepts. In IIT 3.0, however, space must represent both the past and future states, because the theory is built on the notion of $cei$. As a result, there must be two repertoires that give $\phi\textsubscript{max}$: core causes and effects. Since the concepts cover not only the past but also the future repertories, the space where the concepts are represented, and the geometrical structures made from concepts are also doubled. In other words, the multidimensional space and MICS must cover both of the past and future. As the definition of information changes, almost everything in the geometry of IIT appears to be doubled: points, space, and geometrical structures. In this sense, the geometry of IIT seems to be rather complicated. 
\par In a nutshell, the evolution of the geometry of integrated information has two sides. On one hand, it has been simplified in that all the articulated vector analyses for q-arrows are not used anymore. On the other hand, however, it has been complicated in that every aspect of the geometrical approach should be counted twice. We believe that these double aspects of the geometry of IIT are consequences of transitions to other central notions of the theory. 

\section{Theoretical Issues in IIT}
\label{4}
Despite its scientifically interesting prospects, IIT also faces several theoretical problems. These problems concern IIT’s principles, core concepts, and their possible consequences. Despite a few exceptions, many recent literatures are almost focused on particular technical issues\parencites{OizKitzKanai2018,OizHid2018}.\footnote{Rare exceptions are \parencite{Bayne} and \parencite{KroOst2017}. The former provides critical assessments of the axiomatic approach of IIT 3.0. The latter illustrates the important and disturbing conceptual issue of “magic cuts” which can violate IIT’s fundament intuition: “the whole is more than the sum of its parts”.} This focus is fully understandable, since, without overcoming various technical barriers, there would be hardly empirical advances for IIT. However, theoretical problems deserve more attention, as it is theoretical considerations that enable us to judge whether or not the theory is worth pursuing in the first place. Despite such importance, theoretical issues have been largely overlooked in IIT debates, and relatively fewer studies have addressed this topic. Therefore, such theoretical problems IIT require a closer analysis. Of note, while there might be many issues concerning IIT’s theoretical aspects, in the present paper, we focus on three major problems that appear to raise serious questions about the plausibility of the theory. 

\subsection{Sophisticated panpsychism: Unjustified scientific authority}
\label{4.1}
The first issue is that IIT embraces a form of \textit{panpsychism}. Panpsychism has traditionally been ignored as full-fledged mysticism. The view that extremely simple organisms and even seemingly non-living things have ‘a small piece of mind’ sounds counterintuitive enough. IIT, however, admits a variety of examples that could support a sophisticated sort of panpsychism. A representative case is that of photodiodes\parencites{ton2008,IIT3.0}. According to IIT, a photodiode, which is designed to react to various external stimulations only by lighting on and off, is “minimally conscious.” It means that the photodiode has a minimal level of consciousness and a certain quality of experience as well. Nonetheless, the photodiode might be the last one we ascribe experience to. It is difficult to believe that such a simple micro-mechanism could have a certain kind of consciousness. There is another example which appears to be the opposite of the photodiode case. Aaronson\parencite{Aaronson2014a} has clearly shown that, if IIT is right, a lattice constituted by just connecting one kind of simple logic gates over and over could have a high value of $\Phi$. According to Aaronson’s description\parencite{Aaronson2014a}, XOR gates arranged in a 2D square grid would be conscious. Much incredible result is that such increasing of $\Phi$ is proportional to the length and breadth of the grid.\footnote{To be fair, Aaronson’s calculation was based on IIT 2.0 so that it cannot be directly applied to the current version. Unlike IIT 2.0, IIT 3.0 does not require procedure of normalization. Thanks for the reviewer who reminded this point.} Therefore, by a simple recursive procedure of connecting more XOR gates, there could always be a huge physical lattice which is more conscious than a normal human being! Though this is truly unbelievable, IIT clearly allows these examples.
\par If the photodiode refers to the micro-case of panpsychism, the lattice could be its macro-case. The problem is that those simple and non-organic things’ being conscious is so counterintuitive that it would rather be easier to take it as \textit{counterexample} than as evidence. If IIT predicts that those simple systems which are apparently unconscious could be conscious, at least for many, such prediction itself would be enough to present \textit{reductio ad absurdum} against IIT. Hence, the charge of panpsychism should be taken seriously in deciding whether or not IIT is theoretically plausible. If it is certain that no simple object such as a photodiode or a XOR grid is conscious and therefore panpsychism is wrong, IIT must be wrong too.
\par What makes this issue more problematic is that the founder of IIT is seemingly undaunted by those critiques above. IIT does not just allow diverse panpsychistic cases. It actually argues for it, by demonstrating those counterintuitive cases in a detailed manner. Tononi’s reply\parencite{Ton2014Aar} shows his confidence that all those counterintuitive cases are actually the evidence for IIT. Tononi emphasizes that when science and popular intuitions or common-sense conflict with each other, it is always science that takes priority\parencite{Ton2014Aar}. According to Tononi, this is the primary reason why we should count those hard-to-swallow examples as evidence\parencite{Ton2014Aar}. The history of science is full of reversions of commonsense by innovative scientific discoveries. Since IIT is a scientific theory, the fact that IIT produces several counterintuitive predictions cannot be a strong reason to reject it. Rather, in Tononi’s view\parencite{Ton2014Aar}, it is our widely entrenched intuition that must be corrected. Said differently, IIT might be on the edge of “scientific revolution,” and Tononi might be, following Aaronson’s witty phrase\parencite{Aaronson2014b}, “the Copernicus-of-consciousness.” 
\par Nonetheless, Tononi’s reply\parencite{Ton2014Aar} could be objected in several ways. First, it is not obvious at all that IIT is really able to claim its priority over commonsense or intuition. Even if it is true that science tends to override culturally and historically widespread intuitions, the question remains whether IIT has any right to do so. Technically, not all hypotheses of science can have the right to correct popular intuitions. In Kuhnian terms, only the so-called “normal science,” which has successfully secured, well-established methodologies, exemplars, problem-solving procedures, basic beliefs and values shared by members of the scientist society, can argue for its right over commonsense and intuition\parencite{Kuhn1962}. However, it seems undeniably clear that, in the current stage, IIT cannot be such normal science of consciousness. For now, it is nothing more than an interesting working hypothesis that should wait for rigorous examination from the current scientist society. Moreover, as repeatedly pointed out in Section \hyperref[3]{3}, IIT suffers from a number of technical issues preventing empirical experiments and practical applications. No direct evidence has been obtained by empirical studies conducted on real physical systems. Despite the growing body of empirical studies resting on the IIT framework, no IIT theorist has been able to apply the pure IIT 3.0 to neural data such as brain signals.
\par Given its present status in the field, IIT appears to be unfit to serve as a hypothesis of normal science. Rather, IIT is more likely to be something in between “pre-science” and normal science, which might be one possible candidate of “paradigm shift” in the field of consciousness studies. Then, IIT’s panpsychistic predictions cannot be prior to our general intuitions about consciousness. A heavy burden of proof is still on the side of IIT, and our anti-panpsychistic intuition should be taken as default. Aaronson’s comment\parencite{Aaronson2014b} reveals this situation: “The anti-common-sense view gets all its force by \textit{pretending} that we’re in a relatively late stage of research—namely, the stage of taking an agreed-upon scientific definition of consciousness, and applying it to test our intuitions—rather than in an extremely early stage, of agreeing on what the word “consciousness” is even supposed to mean(italics added)”. 
\par The problematic implication of sophisticated panpsychism does not lie only in the conflict with the strong intuitions, which is external to IIT. It also lies in the logical development of the structure of the theory, which is internal to IIT. It is the most original and unique feature of IIT that the theory starts from a number of phenomenological axioms. Yet, the problem is that the axioms are taken for granted in IIT. They are assumed to be self-evident. However, taking something for granted or assuming it to be self-evident is just another way of accepting it as intuitive. In this sense, it is IIT itself that strongly depends on a set of intuitions. IIT is fundamentally \textit{grounded} on several phenomenological intuitions.\footnote{To this matter of grounding IIT, one reviewer has raised an interesting point. The reviewer predicted that “defendants of IIT would argue that the set of axioms is qualitatively different from anti-panpsychist intuitions in that they are not only self-evident but also directly accessible from a first-person perspective”. While this might be true, this reply seems to raise another issue about the direct accessibility of consciousness and its fundament properties, on which phenomenological axioms are about. This ‘direct accessibility from the first-person point of view’ has usually been discussed under the title of \textit{introspection}. In order to claim that introspection lends further support to the phenomenological axioms, one must first prove that such introspection is significantly reliable enough to have some evidential force. However, it is controversial if introspection is significantly reliable; rather, a growing number of empirical studies suggest that introspection is not a reliable source of evidence. Once this point is taken, the alleged qualitative difference between anti-panpsychistic intuition supporting common sense and phenomenological axioms grounding IIT becomes doubtable. Though the reliability of introspection deserves deeper analysis, in the current context, raising doubt against introspection is enough to elaborate our argument by blurring the difference between anti-panpsychistic and phenomenological intuition. For a thorough critical assessment of the reliability of introspection, see \parencite{sch2008} and \parencite{sch2013}. Smithies and Stoljar also present ample philosophical arguments for or against the special nature of introspection\parencite{ss2012}.} Hence, if IIT allows panpsychistic cases and denies opposing intuitions, \textit{a charge of double standards} could be raised. On one hand, IIT strongly holds some intuitions by calling them “phenomenological axioms”. On the other hand, it easily dismisses other intuitions by treating them unscientific commonsense. Nonetheless, how can IIT justify this selective adoption of intuitions? Why does it adopt one group of intuitions but reject another? If axioms of IIT are considered as a significant type of phenomenological intuition concerning what consciousness \textit{is}, anti-panpsychistic intuitions should also be taken to be equally important phenomenological insights about what consciousness \textit{is not}. At least in the current version of IIT, we cannot find any principled reason to take axioms for granted and to reject other intuitions about consciousness. Once IIT wants to deny anti-panpsychistic intuitions as prejudices of scientifically unenlightened laymen, it should do the same thing with its own underlying intuitions. However, what such denial of its own axioms really amounts to is just a self-refutation. Therefore, without providing further reason to take its axioms and ignore anti-panpsychistic intuitions, IIT cannot be free of its charge of double standards of contrasting intuitions. 
\par To sum up, sophisticated panpsychism implied by IIT threatens IIT itself in two ways. First, considering IIT’s premature status, the panpsychistic charge gives a very good reason to defy IIT. As long as no strong evidence is provided, panpsychism alone could suffice not to believe IIT. In addition, it raises the charge of double standards to seemingly equivalently respectable intuitions. Being fundamentally founded by “phenomenological axioms”, it is difficult for IIT to dismiss opposing intuitions. 

\subsection{Fading and dancing qualia: Radical dissociation between experience and cognition}
\label{4.2}
The second issue with IIT is that as Cerullo has pointed out\parencite{Cerullo}, IIT faces the fading and dancing qualia arguments.\footnote{Although Cerullo highlights the point\parencite{Cerullo}, he does not provide a specific description or analysis in his work. By contrast, Shanahan provides a more clear and comprehensive analysis\parencite{shanahan2015}. Both of them concerns upon the problem of fading and dancing qualia anyhow.} Fading and dancing qualia are basically thought experiments designed by David Chalmers\parencite{Chalmers1996}. As suggested by their names, fading qualia describe an imaginable situation where qualia become more and more eroded. Dancing qualia show another scenario that the whole qualia are replaced by totally different qualia. Their purpose is to show that, in our natural world, any attempt to detach experience from the functional organization of a system would face extremely counterintuitive consequences. Despite the richness of detail, in the context of IIT, the relevant point is simple: IIT appears to entail anti-functionalism or anti-computationalism so that it commits to a possibility which fading and dancing qualia rule out. 
\par Fading qualia start with the assumption of the physical system and its functional organization in our world. Since functional organization is a matter of abstraction, it must be fixed how far the organization should be grained. In fading qualia, functional organizations are supposed to be sufficiently fine-grained to fix physical systems’ \textit{behavioral capacities}. Following this assumption, if two physical systems share their functional organization, all their behaviors must be identical. Another assumption is \textit{multiple realizations} without experience. It is assumed that there are multiple kinds of materials in implementing one organization, but only some of them support the phenomenal qualities of experience accompanied by the organization, while others do not. 
\par Now let us imagine that the functional organization of Mary’s brain is realized by neurons. Then, Mary sees a ripe tomato and feels a visually red feeling. In her brain, maybe somewhere in her visual cortex, there is a neural correlate of that red quale. However, something strange happens. The neurons composing her neural correlate of phenomenal redness are now substituted by silicon chips one by one. Given multiple realizations, this replacement must be possible. The crucial point is that, although those chips are perfect functional equivalents of Mary’s neurons, they do not support any quale at all. A natural consequence is that her vividly red experience becomes murkier, and eventually disappears. The problem is that Mary’s functional organization never undergoes any change, despite the gradual qualitative change of her experience. She would still manifest exactly the same bodily and verbal behaviors as before. Moreover, considering that her brain function is perfectly the same, it is reasonable to think that her \textit{cognitive states} are also the same as before. If cognitive states of Mary, such as her judgments or beliefs about experience, do not remain intact and change following the eroding visual experience, such cognitive states would radically come apart from the functional organization of Mary’ brain. Nothing in the functional organization would correspond to the change of cognitive states. Chalmers argues that this kind of dissociation is highly unlikely, by saying “If such a major change in cognitive contents were not mirrored in a change in functional organization, cognition would float free of internal functioning like a \textit{disembodied Cartesian mind}”\parencite[p.258]{Chalmers1996}. This is why he claims that “There is simply no room in the system for any new beliefs to be formed”, “[u]nless one is a dualist of a very strong variety”\parencite[p.258]{Chalmers1996}. As free-floating, disembodied cognitive states are deeply problematic and counterintuitive, it is safe to assume that cognitive states do not suffer any change.\footnote{One of the reviewers advised that there should be more rationales to claim that cognitive states are fixed under the gradual replacement. For more on the debate, see \parencite{Chalmers1996}, p.247-274.} As a result, Mary neither notices nor is aware of anything. This is fading qualia in a nutshell. It seems highly unlikely that such situation could really occur in our world. What is worse is that Mary is perfectly rational and functional in every other aspect, except for her beliefs about her visual experience. She is not pathological or deeply confused. Nevertheless, she suffers somewhat systematic errors concerning her experience. Whenever a substitution occurs, she forms a wrong belief that she is still seeing the red tomato. Clearly, this systematic error of rational subject is hardly acceptable in our natural world. 
\par Dancing qualia is another version of fading qualia. In fading qualia, the phenomenal aspect of the experience gets gradually eroded and ends up to none. In dancing qualia, however, the phenomenal aspect does not totally vanish. Instead, it keeps changing itself. Mary does not suffer the gradual neuron-silicon replacement. Nonetheless, she has a certain neuroprosthetic device, which functions identically to her natural neural correlates of the reddish quale. This time, despite its function, the device does not support the reddish quale. Suppose that it grounds a blue quale instead. And there is a switch that alters Mary’s neural correlate to the device. Then, what would happen if someone turns the switch on? \textit{Ex hypothesi}, Mary’s visual experience will suddenly become blue-like. If the switch turns off, the opposite would happen. Hence, as someone turns the switch on and off, Mary’s visual quale will dance back and forth! The trouble is that Mary would not be able to notice any change in her visual field. Since the device is the perfect functional duplicate of Mary’s neural correlates, the functional organization of her brain remains exactly the same. As in fading qualia, Mary’s cognitive states would be intact, regardless of the change of the phenomenal aspect of her visual experience. If so, Mary would neither notice nor be aware of any change, even if visual qualia are dancing \textit{“in front of her eye”!} For the same reason as in fading qualia, it appears that this consequence must be rejected. 
\par The relevant point in the context of IIT is that IIT essentially allows these implausible cases. Fading and dancing qualia are possible only on the assumption that there could be functionally identical, but phenomenally different systems. In the IIT framework, neurons and silicon chips, neural correlates and the neuroprosthetic device could be such systems. The only way to detour the unwelcomed consequences appears to be denying the possibility of the functionally identical, but phenomenally different systems. IIT, however, does not and even \textit{cannot} deny that possibility. According to IIT 3.0, even if two physical systems perfectly share their functions, they can be different in $\Phi\textsuperscript{max}$ they produce. Considering that maximally integrated conceptual information is experience in IIT, the claim that function and $\Phi\textsuperscript{max}$ can come apart implies \textit{anti-functionalism} or \textit{anti-computationalism} about consciousness. There could be zombie systems which perform exactly the same as conscious systems but do not have any experience at all. This is not just a speculation; indeed, Oizumi, Albantakis, and Tononi design such a zombie system and demonstrate how it works\parencite[see][Figure 21]{IIT3.0}. If a zombie system is possible, there is no reason not to believe silicon chips in fading qualia or neuroprosthetics in dancing qualia. Then, IIT should accept those unacceptable consequences anyway. 
\par It is IIT’s anti-functionalism that opens the door to fading and dancing qualia. In front of the implausible results of fading and dancing qualia, there are only two logical ways for IIT to reply: to dodge the bullet or to bite it. Nonetheless, none of the two appears to be available without significant revisions of the theory. On one hand, if IIT wants to dodge the bullet, it must show how Mary could notice the change in her visual experience, even if her brain functions remain exactly the same. It is highly likely that, if there is no difference in the brain functions, the same will apply to information processing. In IIT as a paradigm of cognitive science and artificial intelligence, it is widely accepted that, in order to notice or be aware of something, there should be corresponding activities of information processing. However, by the assumption of functional identity, Mary cannot have any new information processing corresponding to the change of quale. Then, how can Mary notice or be aware of the experiential change? On the other hand, if IIT tries to bite the bullet, all the debates concerning the charge of double standards resurface again. IIT cannot merely say “Though being counterintuitive, it’s true nonetheless”. IIT is scientifically so premature that it is not in a position to simply override strong intuitions in the name of science. Furthermore, since IIT itself takes some intuitions as primitive, it cannot easily dismiss other intuition as ungrounded. In one way or another, it seems difficult for IIT to defy the intuition that the \textit{radical dissociation between cognition and experience} is impossible. In one way or another, IIT can neither dodge, nor bite the bullet of fading and dancing qualia. 
\par All in all, IIT cannot deal with fading and dancing qualia. Holding anti-functionalism about consciousness, IIT does not have theoretical resources to explain how the system which functionally remains identical could notice its phenomenal changes. On the other hand, accepting the possibility of unnoticeable phenomenal change is extremely counterintuitive to that, if IIT allows such notion, many would reject IIT. As in the panpsychism debate, due to its dependence on intuitions, IIT cannot merely dismiss the intuition that a rational and functioning system must be able to be aware of its own experiential changes. Anyway, IIT faces serious troubles. 

\subsection{The paradox of certainty: Loss of certainty undercuts existence}
\label{4.3}
In Section \hyperref[4.2]{4.2}, we argued that, although the empirical possibility of radical experience-cognition dissociation causes a serious counterintuitive consequence, IIT cannot dodge this consequence. In this section, we attempt to show that such radical experience-cognition dissociation causes another problem:  \textit{the loss of certainty about consciousness}. We believe that this loss of certainty can undercut the very foundation of IIT:  \textit{the existence of consciousness}. 
\par We, or at least many of us, appear to be \textit{certain} about our consciousness. Our own consciousness might be the only thing we can be certain about. However, the argument from fading and dancing qualia shows that our phenomenal beliefs or judgments can be detached from our consciousness even when we are fully alert and attended. If this is the case, \textit{we ourselves} might be suffering fading and dancing qualia as well. That is, we might be like Mary who cannot be aware of the absence of her own visual consciousness. If so, even if we strongly believe or take for granted that we are conscious here and now, it is possible that we are not. As Descartes doubted, an omnipotent demon might manipulate our perceptual experience to make us believe the existence of the external world, even if there is no such world. Similarly, something might control our cognitive system to make us believe the existence of our experience, even if there is no such thing as experience at all. Then, how can we be so sure about that we are conscious here and now? In other words, is there any guarantee that we are not \textit{deluded zombies who think that they are conscious} if experience and cognition about the experience can come radically apart? It is clear that the radical experience-cognition dissociation deprives us of the certainty of consciousness. And if IIT allows the dissociation, it cannot secure the certainty of consciousness. 
\par Some might deny the certainty of consciousness. Although the certainty of our own experience appears to be the last thing we can deny, whether or not we are really certain about our experience is surely debatable. Nevertheless, it appears that IIT cannot easily deny the certainty of consciousness, because the theory appears to be \textit{grounded} in it: the first phenomenological axiom states that consciousness exists. Furthermore, this existence of conscious experience is supposed to be certain. Indeed, it is clearly argued that consciousness is certain when Tononi paraphrases\parencite[p.296]{ton2012} Descartes’ \textit{cogito ergo sum}: “I experience therefore I am”. The very starting point of IIT, the existence axiom, necessarily requires the certainty of consciousness. If we are not certain about our own consciousness, why should we struggle for a scientific theory of consciousness? 
\par Therefore, the possibility of the radical experience-cognition dissociation provides a somewhat delightful and disturbing paradox against IIT: If IIT is true, radical experience-cognition dissociation is actually possible. If so, we cannot be certain about our own consciousness. If we cannot be certain about our own consciousness, IIT cannot get off the ground. Therefore, if IIT is true, there is no reason to suppose that it is true. We call this argument \textit{the paradox of certainty}. IIT appears to simultaneously require and reject the certainty of consciousness. 
\par It seems that the only possible reply from IIT would be denying the empirical possibility of the radical experience-cognition dissociation. However, as we have seen in Section \hyperref[4.2]{4.2}, the problem is that, at least in the current version of the theory, it is difficult to find any rationale for such denial. Considering the fact that IIT actually argues for the functional zombie system, it is doubtable that IIT can deny such possibility. In fact, we cannot find any consideration about how experience affects beliefs or judgments, and vice versa in IIT. While IIT appears to have a great deal with how experience is generated from its physical substrate, it does not provide much insight into how the subject can be aware of that generated experience. Said differently, IIT is blind to the question of how we can secure \textit{self-knowledge} or \textit{metacognition} about our own experience. This is the topic of the last section of this paper. 

\subsection{Metacognitive accessibility: Missing link in IIT}
\label{4.4}
What is the main source of the theoretical problems mentioned thus far? We think the culprit here is disregarding cognitive aspect of consciousness.\footnote{Cerullo makes a similar point\parencite{Cerullo}. After distinguishing \textit{incognitive} and \textit{cognitive} consciousness, he argues that IIT only deals with incognitive one, which is tantamount to consciousness without subject. } In IIT, the explanation of how the experience could be cognitively accessed by a subject is totally absent. IIT never takes account of \textit{metacognition} in explaining consciousness, and we believe that it is this neglect of metacognition that generates all theoretical problems IIT faces. 
\par Due to its ignorance of metacognition of consciousness, IIT can ascribe consciousness to simple systems lacking metacognitive mechanisms, such as photodiodes or logic grids. Though photodiodes and logic grids produce integrated information, it is highly unlikely that these simple physical systems are equipped with metacognitive mechanisms. Given that they lack metacognition, those systems do not, and even cannot, have cognitive access to integrated information of their own. There is no photodiodes and logic grids’ metacognition of their integrated information. Under the IIT framework, this metacognitive inaccessibility implies that photodiodes and logic grids cannot know or be aware of their own consciousness. While they are conscious, they cannot know that they are conscious! However, this lack of metacognition and its strange consequence do not prevent IIT to ascribe consciousness to simple systems, as it does not concern metacognitive access to consciousness at all. 
\par Furthermore, since IIT appears to neglect how metacognition and experience could be associated, it allows the radical dissociation between metacognition and experience, which is shown by fading and dancing qualia and ultimately results in the paradox of certainty. In fading and dancing qualia, unlike in the panpsychistic cases, the system has metacognitive access to integrated information it produces. That is, Mary has a metacognitive belief about her visual experience. The problem is that her metacognitive access systematically produces wrong beliefs about her own experience. In fading qualia, Mary is usually right about what she sees. However, as soon as the process of neuron-to-silicon replacement begins, Mary starts to have wrong beliefs about what she sees. In dancing qualia, whenever the switch turns on, Mary becomes wrong about her visual experience. In both cases, Mary’s being wrong is very systematic in that it strongly correlates with the replacement. Mary’s systematically being wrong indicates that her metacognitive access to her visual experience systemically results in wrong beliefs. However, since there is no consideration about how the system metacognitively accesses its own experience in IIT, it cannot help but allow the absurdities of fading and dancing qualia. In addition, once the radical experience-cognition dissociation is admitted as possible, there appears to be no way to eschew the paradox of certainty. 
\par Given the tight relationship between experience and cognitive access, IIT’s neglect of metacognition is somewhat surprising. Phenomenologically, there appears to be a close, even constitutive relation between metacognition and experience. Despite philosophical debates surrounding the distinction between phenomenal vs. access consciousness\parencites{Block1995,Block2007}, we believe that there could be experience without actual metacognitive access. Nevertheless, this does not mean that there could be an experience that cannot be metacognitively accessible. It sounds absurd and even unintelligible that a conscious experience is absolutely out of our range of metacognition. Such experience must be a \textit{conscious experience we cannot be conscious of}, which is unconscious by its nature. Hence, it appears that metacognitive \textit{accessibility}, not actual metacognitive \textit{access}, is necessarily involved in having consciousness. That is, metacognitive accessibility is a necessary condition for something to be a conscious experience.\footnote{For a similar point, see Chalmers\parencite{Chalmers1997} who argues against Block\parencite{Block1995} that, even when there is \textit{phenomenal consciousness}(P-con) without \textit{access consciousness}(A-con), it does not mean that there is not accessible consciousness. According to Chalmers, once A-con is defined in terms of availability for global control, P-con always goes along with A-con\parencite{Chalmers1997}. Since global availability requires only accessibility, the original notion of A-con should be modified from access consciousness to accessible consciousness. Our suggestion here could be taken as claiming that, if an experience is phenomenally conscious, it must be accessibly conscious. It is worth noting that this transition from access to accessibility is what distinguishes Chalmers\parencite{Chalmers1997} and us from those who follow \textit{Higher Order Theory of consciousness}(HOT)\parencites{rosenthal1986,rosenthal2005}. In HOT, for a mental state to be conscious, it must be actually accessed by a higher order state. Our suggestion, however, does not demand actual higher order, metacognitive access. All that required is that the state must be metacognitively accessible. No actual higher order state needs to be there.}
\par Therefore, we argue that any scientific theory of consciousness must take account of the metacognitive accessibility of consciousness. However, no matter which version it may take, IIT does not seem to consider why and how metacognitive accessibility must be taken into account when it comes to explaining conscious experience. Accordingly, we strongly suggest that the first step to deal with the theoretical problems mentioned so far is introducing metacognitive accessibility in the IIT framework. Phenomenological axioms, ontological postulates, and mathematical models of IIT should be revised in order to reflect the necessary connection between metacognitive accessibility and consciousness. Once we can successfully assimilate metacognition into IIT, we could have a better version of the theory, which would deserve to be called ‘IIT 4.0.’ 

\section{Conclusion}
\label{5}
IIT has been a center of the debate surrounding the science of consciousness. Many of those who are engaged in the field displayed interest in the theory, and some raised serious doubts and criticisms. It is worth to assess what IIT is about and why it is controversial. In this paper, we have critically examined the theoretical evolution and related issues of IIT. We have introduced basic concepts, which might be considered as the core of IIT. Both IIT’s explanatory power and limits appear to be already embedded in its core concepts. We have also described how the theory has been updated throughout the last decade. In some aspects, those major transitions can be thought as a progress. However, in other aspects, some of the issues were worsened, and even new problems emerged. Specifically, the principled part of the framework of IIT, its phenomenological axioms, and ontological postulates raise serious questions about the scientific status of the theory, the possibility of radical dissociation between experience and cognition, and the logical structure of the theory. We have suggested that focusing on our ability to access our own experience through metacognition might be one way to deal with these theoretical issues. The cognitive relationship between metacognition and consciousness might push IIT one step forward in becoming the science of consciousness. 

\section*{Author Contribution}
HP wrote Section 2 and 3; KM wrote Section 1, 4 and 5; All authors reviewed the manuscript; HP documented the manuscript in \LaTeX.

\printbibliography

\end{document}